# Unlocking WebRTC for End User Driven Innovation

Kundan Singh

Intencity Cloud Technologies
San Francisco, CA, USA
kundan10@gmail.com

**ABSTRACT**

We present a software architecture to enable end user driven innovation of web multimedia communication applications. *RTC Helper* is a simple and easy-to-use software that can intercept WebRTC (web real-time communication) and related APIs in the browser, and change the behavior of web apps in real-time. Such customization can even be driven by the end user on third-party web apps using our flexible and general purpose browser extension. It also facilitates rapid prototyping of ideas by web developers in their existing web apps without having to rebuild or redeploy after every change. It has more than ten customization categories, and over a hundred built-in examples covering a wide range of novel use cases in web-based audio/video communication.

## 1. INTRODUCTION

The inclusion of web real-time communication (WebRTC) API [1][2] in popular web browsers has prompted several existing and new application service providers to embrace this technology. However, open innovation is cripled by vendor lockins and walled garden approaches to such web applications [3][4][5]. Moreover, WebRTC has intentionally left out the signaling or control pane specification. This further promotes creating islands of apps, where the users of one *app* or website cannot easily communicate with those on another, even though the underlying tool or the browser is capable of facilitating such communication. The result is that the top few popular communication and conferencing apps dictate the complete user experience for a vast majority of end users.

This tight coupling between the *app* and *service* by each vendor hinders innovation. For instance, if a user signs up with a conferencing service, she must also use the client app by the same vendor, and cannot easily use other apps for features like note taking or virtual background. Once the vendor successfully locks in enough users, there is little motivation to innovate beyond the features-list of the top few products in that market. The continuous innovation in the space remains far fetched if the customer is locked into a single platform or app. For example, it is hard to use auto-framing in webcam capture, or a third-party quality monitoring tool, unless that vendor supports or allows it.

Although an installed app cannot be forced to such demands, the web architecture is more flexible and open. Web APIs separate the web apps from the browser user agent. Intercepting and modifying such web APIs have been done for a long time, such as for ad-blockers, workaround for older browsers, for cross browser support, or monkey patching of web API by various JavaScript frameworks and tools. In the context of WebRTC, this enables us to do some app mashups at the client end, and allows a number of end-user driven innovative use cases. For instance, one can inject a new video background detection on a third-party web app, or capture and send the call quality metrics to a monitoring service independent of the video conferencing app. This can be done by the developer by changing the web app's source code, as well as by the end user via a browser extension.

We present *RTC Helper* [11] – a simple and easy-to-use software that intercepts WebRTC and related APIs between the app and the browser. It then allows injecting code snippets in pure JavaScript to customize the behavior. Our objectives are: (1) to allow end-users to customize their multimedia communication experience on any such web app, and (2) to allow web developers to quickly create implementations of innovative ideas and emerging features on top of WebRTC and related APIs.

We list motivational use cases in Section 2, and provide background and related work in Section 3. Section 4 shows the software architecture. Implementation overview is in Section 5, and security discussion in Section 6. Finally, we conclude in Section 7. We include additional implemented use cases and details in Appendix A and B.

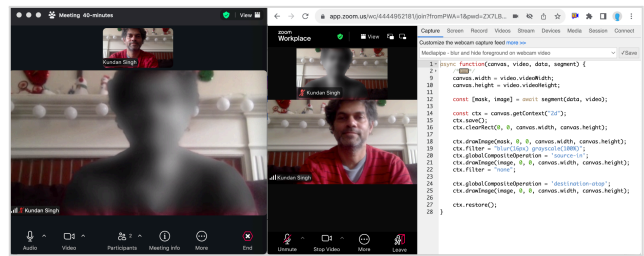

**Fig. 1** Foreground blur using RTC Helper on a Zoom video call.

## 2. MOTIVATIONAL USE CASES

Following use cases or scenarios illustrate end-point driven innovation, but are often missing in web conferencing apps.

A popular feature in video conferencing is background blur, where body segmentation is used to detect foreground. The same algorithm can also be used for blurring the foreground face and body for privacy, while showing the unaltered background (Fig.1). But this is not available in top video conferencing products because of its niche demand.

A video conferencing participant on a webcam would like to temporarily replace her video with her mobile device's camera feed to show something around. This is not easy to do in existing products that require her to join as another participant from her device to accomplish the task.

A salesperson sharing her product presentation would like to include her company logo, watermark, or confidentiality label on top of her screen share feed. She would likely need to edit her presentation material to do that, instead of a live quick fix on her screen share video feed.

A presenter would like to include her webcam video in a small circle on a corner of her shared slide deck, instead of being displayed in a big video box of the app layout, taking up unnecessary screen space (Fig. 2).

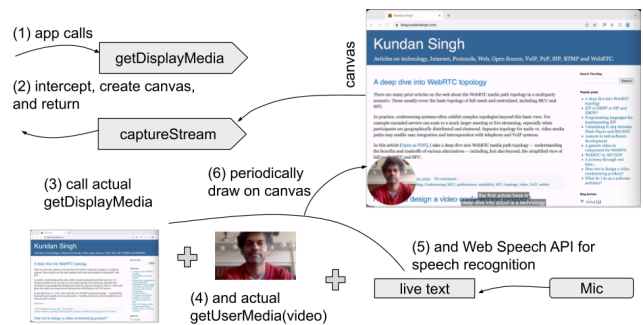

**Fig. 2** Screen or window share with auto-injected presenter video, and captions transcribed from presenter's microphone.

Many existing video conferencing apps allow only a single content share. If the user would like to share two windows, she could likely share her whole screen, and minimize all other windows. Could she instead select the two windows, create a single video feed that combines the two, and share that in her existing app?

Ability to record a video conversation is often available in conferencing apps, but is controlled by the vendor, e.g., in what video layout or view, for how long, and at what cost. It would be nice if the end user could record the conversation easily from her perspective, within the legal boundary.

Video conferencing apps often display all webcam and content share videos in the same window. For users with dual monitors, it would be nice to separate out the content video to a separate window on the second monitor. The presenter may prefer to minimize the conferencing app, but still be able to view other participants' videos albeit in a smaller size.

A user has a bad microphone device that she cannot quickly detach from her machine. But this device causes problems with her video conferencing app. Could she hide that device from the particular app?

A similar user has a bad camera device that crashes every time it is captured at 1080p by the app. Could she force the app to capture it at a different or lower resolution to avoid the crash?

A user joins a video conference on her mobile data plan, and would like to restrict the bandwidth used by the video codec. But her video conference app does not have any such option. She will likely disable her video altogether.

An organization with strict accountability would like all her employees to avoid peer-to-peer media paths of WebRTC, and to always go through the approved media relay installed by the organization. Can it force this on the third-party web app that the employees use?

A similar privacy conscious organization would like to embrace WebRTC, but is nervous about the data channel's opaqueness. Can it disable the data channel on third-party web apps used from within its network?

A web app developer would like to test her client software under failure conditions. How could she emulate delays, random disconnects, or other errors quickly without having to change the app, rebuild, or redeploy.

Those are some of the use cases that are enabled by our software. There are many other customization functions pre-built in our software, and described in Appendix A. The end-user or web developer can also quickly create more customization functions for many other use cases.

## 3. BACKGROUND AND RELATED WORK

WebRTC enables a web page to establish a peer connection between two browsers and transport captured media from one to another [1]. Fig. 3 shows various important elements and constructs in an asymmetric call with audio and video in one direction and audio-only in the reverse.

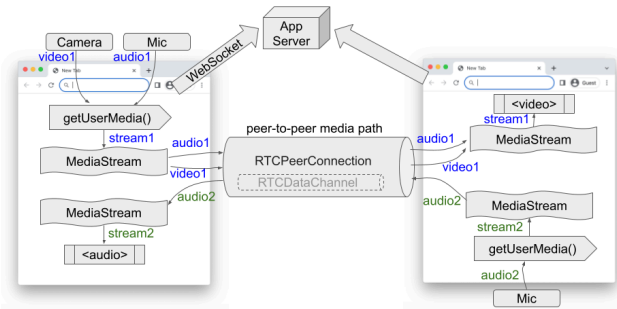

**Fig. 3** WebRTC API elements in an example session.

The WebRTC API defines JavaScript (JS) objects and functions, e.g., getUserMedia and RTCPeerConnection. A notification or app service is used to exchange certain signaling and connectivity data between the browsers. This is often controlled by the web app vendor, and uses a client-server session on WebSocket or HTTP long polling.

The objects and functions of the API can be intercepted and replaced with some custom processing to provide advanced features as shown in Fig. 4.

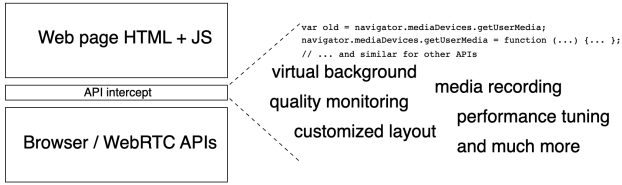

**Fig. 4** Basic architecture of any API intercept layer.

Many successful projects have used this model, such as popular ad-blockers, or JS frameworks. For WebRTC-based apps, this has been done for numerous use cases, such as to prevent local IP address leak, to enforce enterprise policy on peer-to-peer traffic, to capture and send logs and quality metrics to a monitoring service, to prompt the end user to grant individual feature access, to intercept media path for testing and debugging, or to redirect media path in a virtual desktop environment [6][7][8][9][10]. The existing effort has largely focussed on testing, debugging, or policy enforcement of the WebRTC apps, either being agnostic to or reinforcing any vendor lock-ins of such apps.

The novelty of our effort is in unlocking such apps to the end users, not just for testing and debugging, but also for improving their audio/video communication experience by bringing unique or tailored features as per individual needs. Some specific use cases enabled by our software are already present in some existing systems, or have been explored in the past. For example, video pop-out or picture-in-picture layouts are available in popular conference apps like Zoom and Teams; Apple's continuity camera allows us to use the iPhone camera as the webcam feed on Mac; and XSplit VCam and recently, macOS 15+, have an app independent virtual background at the camera capture stage itself.

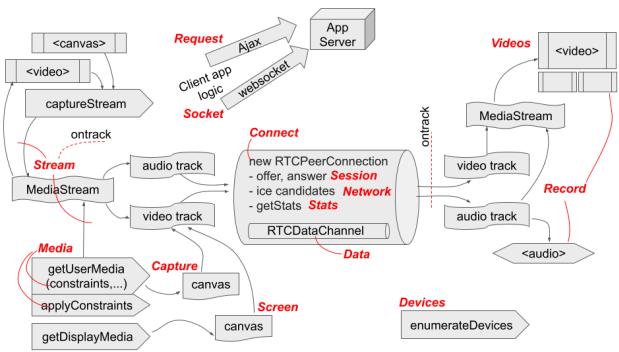

**Fig. 5** Intercept of APIs in various customization categories.

Our approach is different as it provides a generic tool for web apps that covers all these and many more use cases. Furthermore, our developer friendly interface also enables web developers to quickly test out such features live in production systems. We believe our approach incentivizes more innovation in this space, by opening up many client side features to endpoint and end user driven customization, instead of being locked by the web app provider's choices or restrictions.

## 4. SOFTWARE ARCHITECTURE

In our *RTC Helper* software, instead of a developer oriented generic intercept layer that works for API logging, testing or debugging use cases, we identify user friendly scenarios from the motivating examples, and split them into fourteen distinct categories (Fig. 5). Then we provide an intercept layer for each category. The same API element or construct may belong to multiple categories for a cleaner abstraction.

These customization categories are summarized in Table I. More categories that are unrelated to the web APIs, such as Controls, Security, and CPU, are not listed below.

**Table I** Summary of customization categories.

| | |
|---|---|
| Capture | Create an image for each frame in a canvas, which is used as webcam video by the app. |
| Screen | Create an image for each frame in a canvas, which is used as screen or window share. |
| Record | Create a video recording combining multiple video elements, and mixed audio. |
| Videos | Modify the audio and video elements on the web page, especially for layout or display. |
| Stream | Modify any media stream captured locally, or received from remote on a peer connection. |
| Devices | Modify the list of devices available to the app. |
| Media | Modify constraints of a captured media stream. |
| Session | Modify session description containing codec parameters in a peer connection. |
| Connect | Modify peer connection configuration during creation such as for media relays. |
| Network | Modify transport parameters such as ICE candidates in a peer connection. |
| Stats | Capture or modify quality metrics and related statistics in a peer connection. |
| Data | Modify the behavior of a data channel including data sent and received. |
| Socket | Modify the behavior of WebSocket including data sent and received. |
| Request | Modify the Ajax request and response API. |

Each category defines its function signature, and its allowed parameters. The end user can supply a function in each category. The software invokes such installed functions with the parameters that are asked for at various stages of WebRTC and related APIs. For instance, the Media category that deals with microphone or webcam constraints, defines the function to take a constraints object, which can be altered by the installed function, e.g., to remove audio, or to limit the maximum framerate. Similarly, the Session category that deals with any session description, defines the function to take the associated peer connection and session description objects, which can be altered, e.g., to prefer G.711 over Opus as audio codec, or to limit the bandwidth.

© 2025, Kundan Singh　　　　3

These parameters of various category functions are detailed in Appendix B.

## 4.1 WebRTC API Intercept

Note that our software does not intercept an API unless a category that deals with that API has a custom function installed in the software by the end user or web developer.

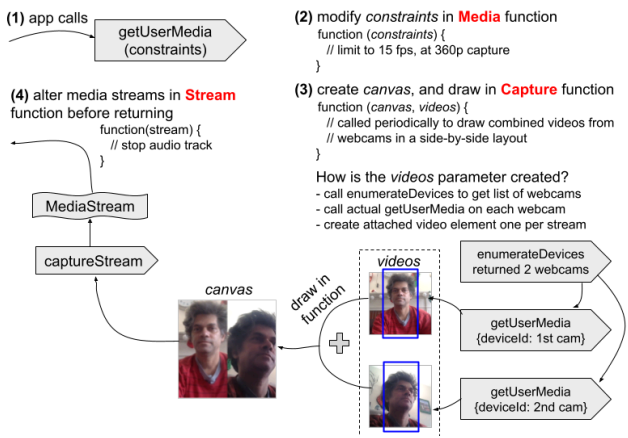

**Fig. 6** Example intercept of getUserMedia in various categories.

When the web app invokes getUserMedia to capture from the local camera and microphone (see Fig. 6), the software first checks if Media is customized, and calls that function to modify any desired media constraints.

If the Capture function is installed, it uses an internal canvas stream to replace the camera track, while leaving the microphone as is. Based on the desired framerate, it then periodically calls the installed Capture function to draw on the canvas. The function can ask for one or all webcam videos if the user has multiple cameras, or even a screen share video. This is done by using named parameters in the function definition, e.g., video, videos or screen. This allows mixing multiple webcams into a single video track, or combining screen share with webcam in a single video track. The web app is given the stream captured from the canvas. Stopping the returned stream by the app causes the cleanup of underlying canvas and related constructs.

If the Stream function is installed, then that is invoked before returning the stream to the web app. This function may completely replace the audio or video track, e.g., from a third-party video feed, which is then given to the app as a captured stream. The Stream function is also invoked when the web app calls captureStream on any media or canvas element, or when the track event is received on a peer connection object. This allows customizing any captured or received media streams.

When the web app invokes getDisplayMedia for screen or window share, the behavior is similar – the software checks for Screen function, and if present, returns a canvas stream, while allowing the function to periodically draw on the canvas for each frame (see Fig. 2). The function can ask for one or all webcam videos as well as one or more screen or window shares. This allows combining multiple window share streams in a single video track given to the app. It also alters the logic as described before if Stream is installed.

The Videos function is called for every audio and video element on the web page, on creation, destruction, state change, or mouse click. This allows the function to change the property or layout such as to show controls or move it.

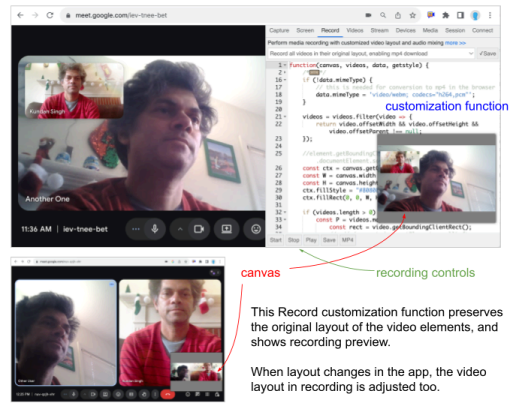

**Fig. 7** Local recording using RTC Helper on Google Meet.

The Record function, if defined, is invoked periodically depending on the desired framerate, to draw on a canvas for recorded video content (see Fig. 7). It can ask for all the video elements displayed on the page. It implicitly mixes and records audio, but the function can configure certain options, such as to disable audio in recording, to use audio track from peer connections instead of audio or video elements, and to use the microphone captured audio track too, if it is not attached to any audio element. The end user can start/stop the recording independent of the web app, and can replay or download the recorded video file.

The Devices function, if installed, can modify the media devices list in the app, when it calls enumerateDevices.

When the app creates a new RTCPeerConnection object, and any of Connect, Session, Network, Stats or Data is installed, then a proxy object is created and given to the app, so that the software can intercept its constructor, other methods and event handlers. Afterwards, depending on which category function is installed, and what function or event handler on the proxy object is invoked or installed, the category function is called. For example, the installed Data function is called for createDataChannel method as well as the datachannel event on the peer connection object. It is also called for the send method and the message event on the underlying RTCDataChannel object. This allows the Data function to customize all parts of the data channel feature. Similarly, the Connect, Session and Network functions deal with one aspect each of the peer connection.

The Stats category, on the other hand, is slightly different. If the function is installed, the software calls getStats on the peer connection periodically, and delivers the statistics to the function. This can be used to locally save or to send the statistics to a monitoring service. To calculate the quality metrics, and to display them, the software includes helper methods to query or combine the statistics data, and to plot any line graphs to display to the end user.

### 4.2 Network API Intercept

A Request function modifies fetch and XMLHTTPRequest in JavaScript, both request and response. This can be used to emulate API responses for testing and debugging. The Socket category allows modifying the WebSocket behavior, including its connection attempt, maintenance, and data exchange. This can be used to intercept and emulate signaling messages in many WebRTC apps.

These functions can only modify the data that are allowed and visible in the web app's JavaScript. However, certain information such as security related HTTP headers are not visible there, but are available in the browser extensions.

The Security category is useful for modifying headers in any HTTP request or response using the underlying webRequest API. This allows altering or relaxing certain headers such as x-frame-origin or content-security-policy (CSP), e.g., if another category uses a third-party script that is not allowed by the web app's default CSP. Unfortunately, Chrome disabled support for the underlying construct in newer browser extensions (manifest version 3). Hence, this category is no longer applicable on Chrome, but can still be used in other Chromium based browsers (e.g., Brave) that still support the older extensions (manifest version 2).

### 4.3 Shared Variables

The CPU category is useful in continuously monitoring the CPU load on the machine, and to take certain actions when a threshold is reached. The underlying CPU monitoring logic runs in the browser extension, and cannot run in the web app. It can trigger events via shared variables that can be received and acted upon by functions in other categories.

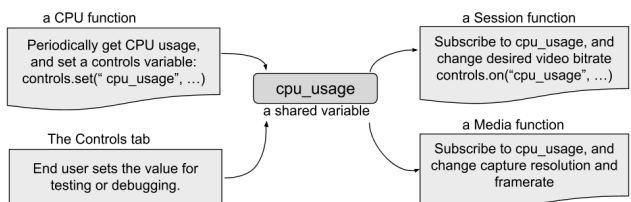

**Fig. 8** An example shared variable used by multiple functions.

Our software allows different category functions to share data using shared variables called controls. Such variables have primitive types such as string, boolean, or number. They are read/written or subscribed to programmatically by a category function, or in the user interface by the end user. Subscribing to the variable allows receiving an event when the variable's value is updated, or it is deleted (Fig. 8). It also supports generic event listening and triggering.

This shared variable feature allows interaction among the functions of different categories, e.g., the CPU function can trigger the load value, and the Media function can adjust the constraints to reduce the video resolution to mitigate the load. The end user can also trigger or edit the controls, and inject the user interaction or input in the category function. For example, users can emulate multiple button clicks on the web app via a single trigger in controls, or set the file path of the logo overlaid on screen capture in a live call.

## 5. IMPLEMENTATION

*RTC Helper* [11] is implemented in JavaScript, HTML and CSS. It has about 13,000 source lines of code, not including the external dependencies such as the code editor, ffmpeg, or mediapipe. The three main parts are the *content script* which is injected in the web page, the pre-built collection of customization *functions*, and the *options panel* that shows the user interface to select or edit such functions, and to allow changing the shared variables and settings.

### 5.1 Install Options

There are three ways or *options* to install this software: (1) a browser extension by the end user, (2) complete integration in a web app by the web developer, or (3) integration of selected features by the web developer. The content script is almost the same in these options, but the options panel is opened differently in the first two, and is missing in the last.

***Option 1***: Any end user can install and enable this to test, debug and customize multimedia communication on any website using a Chromium-based browser. The extension icon appears as a browser action next to the search or address bar.

When clicked, it opens the options panel in a new browser tab. Alternatively, if configured so, it opens as a side or bottom panel within the visited web page (see Fig. 9). When opened in a new tab, the left side allows selecting the website domain to customize, and the right side includes the tabs for the customization categories. When opened as a side panel, the website selection is hidden, and the tabs shown are tied to the opened website's domain. The options panel mimics the visual style of Chrome's developer tools.

The extension can intercept WebRTC and related APIs on all the web pages the user visits in that browser. If the extension detects one of getUserMedia, getDisplaymedia or enumerateDevices called by the visited webpage, it notifies the end user, prompting her to customize the web page, if needed. If the user clicks and chooses to customize, it opens the options panel. The user can click on the browser action icon of the extension on any web page to open the options panel for that website.

Various tabs in the options panel allow the end user to select a pre-built function in that category, or to edit existing or create a new function. The last few tabs are for user controls, settings and help. If some category functions are selected, installed or defined, the WebRTC and related APIs are intercepted and modified as described earlier.

*Option 2*: A web developer can integrate the complete software in her web app, and make it available to all her users. This is done by simply including a couple of <script> elements to load the content script and the options panel of the software. The URL parameters on the script source can be used to configure the panel settings, e.g., to disallow change of some options, or to set their default values.

The user experience and the features are largely similar in the first two install options. In particular, the options panel behaves similarly in the two cases, with two main differences. First, the panel is tied to the specific web app in this option, and there is no website selection. Second, instead of clicking on the extension icon to launch the options panel, the user clicks on a hidden button on the top-right corner of the web page, or uses a keyboard shortcut (Meta+Shift+J/Mac or Ctrl+Shift+J/Windows) to open or close the options panel.

*Option 3*: A web developer can partially integrate a subset of the features in her web app. This is done by including a single <script> element loading the content script. Instead of giving full control of all customization functions to the end user, the developer controls which function is applied and when. The options panel or the hidden button are not included in this mode. Instead, the developer selects or defines the specific functions, and programmatically installs them by invoking the content script directly.

This option allows the web developer to cherry pick specific customization functions such as foreground blur or media recording, and control them in her app logic.

## 5.2 Software Settings

The settings tab is shown with the install option 1 or 2, but not with 3, and allows customizing certain behavior of the tool itself (Fig. 9a). The settings are summarized below.

**theme**: controls the color scheme; allowed values are light or dark, and defaults to using the underlying browser or system settings.

**mode**: controls how the options panel is opened, e.g., as a separate tab or window, or inline in the web page, and if inline, whether as a floating panel which hides the web page content, or which resizes the page, and whether it appears on the right (default), left or bottom side of the web page.

**tabslist**: controls which tabs in the panel are shown, and in what order.

**tabswrap**: controls whether the displayed tabs in the header for navigation are wrapped or scrolled (default).

**fontsize**: controls the font size as small (default), medium or large.

**editor**: controls the behavior and appearance of the code editor, and includes a comma separated list of attributes that are applied to the popular Ace code editor that we use.

**scripts**: controls whether the end user can edit the code in various tabs, and save the code as a named script for later use. Without named scripts, only one active script is saved per tab. With named scripts, the user can experiment with multiple scripts and quickly switch among them.

**urls**: a list of https URLs, one per line, to download the files containing additional functions for various tabs. These functions appear in selection dropdowns in various tabs as applicable. This allows extending the software directly by the end user.

**strict**: controls whether the customization functions are run in strict mode or not. Using the strict mode prevents access to globals and other data in the functions used in various tabs. This is useful if the user does not trust the source of those functions, or if the function definitions are minified to hide its behavior. The strict mode is not needed if the functions are written by trusted sources, or if the user can visually inspect them and find no malicious usage.

**savestats**: to send and persistently save the statistics data, a

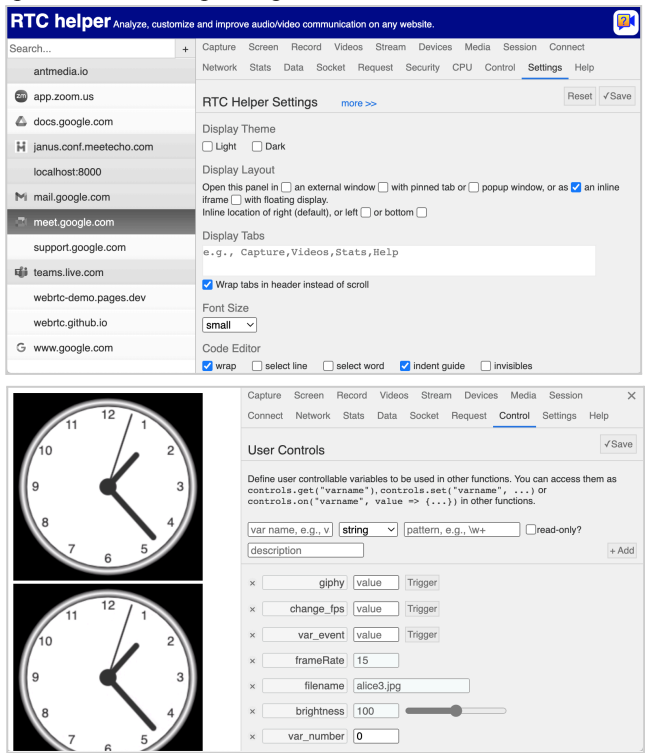

**Fig. 9** The options panel is (a) opened in a new browser tab, showing the Settings tab, or (b) inline on the right side within our test web page with customized capture, showing the Controls tab.

  

Stats function can use an external service. This controls where the data is written to and where it is accessed from.

**ffmpeg**: to convert the recorded video to mp4 format, a specific ffmpeg command is run, such as to transcode video from the default h264 to mpeg4 encoding. Transcoding is slow, and takes up processing resources. The user can change the command to avoid transcoding, if her video player is able to play h264 encoded mp4 format.

Some tabs such as Record and Controls have their own user interface, e.g., to start, stop or download recording, or to trigger an event.

### 5.3 Function Signature and Parameters

A customization function supplied or selected by the end user is just a regular and unnamed (anonymous) JavaScript function, which is defined to take some named parameters. A Capture function snippet taking three named parameters is shown below.

```
function (canvas, video, data) {
    … // draw on canvas using video content
}
```

Some categories allow using an asynchronous (async) function, because the underlying web APIs are also async. However, some categories do not allow async functions, such as Stream, Connect, Network, Data, CPU, Security. These may still use asynchronous code inside the function definition, e.g., using new Promise.

The software passes the right objects as parameter values based on their names, e.g., if the Capture function uses the video parameter, then a single user selected webcam is captured, but if videos is used, then the software captures all the attached webcams, and passes a list of video objects, one from each camera, as shown in Fig. 6.

Appendix B describes the parameters for each category.

### 5.4 Case for Rapid Prototyping

A core objective of this software is to allow web developers to quickly create proof-of-concept (PoC) implementations of new ideas. The ability to write code and immediately see it in action is already facilitated by this software using real-time code injection, and a code editor for functions in various categories.

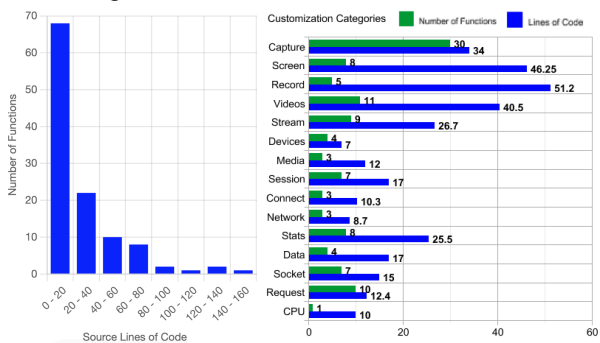

**Fig. 10** (a) Distribution of functions by their sizes in lines of code, and (b) count and average size of functions by their categories.

Moreover, an individual function is usually small within a few tens or hundreds lines of code. All of the hundred or so pre-built customization functions included in the software are only about 3,000 source lines of code in all. Fig. 10(a) shows that most of these functions are very small, and only a handful are more than 100 lines of code.

A further breakdown of the code complexity in terms of the number of functions and the average lines of code in each category is in Fig. 10(b). We include a number of pre-built functions in the Capture category. A number of functions in the Videos category are related to opening and managing new windows, and in Stats are about manipulating complex statistics data to get simple metrics.

Although, the number of lines of code is not a true measure of the code complexity, it gives a good idea about how easy it is to do rapid prototyping of various use cases in this software. It also shows that intercepting and modifying peer connections and related APIs is a lot simpler than creating or manipulating video frames or video elements.

### 5.5 Limitations

The software does not work for web apps accessed from mobile browsers that lack support for browser extensions, or any third-party installed apps. There is no way to intercept the web APIs in those cases. But it can still work for developer driven customization (install options 2 or 3) on installed apps created using web app frameworks such as Electron or Apache Cordova.

The customization function must be in pure JavaScript, and cannot easily use advanced JS frameworks such as Angular, Vue, or React, especially in the strict mode. This may be challenging for some web developers.

The content-security-policy (CSP) of a web app is often delivered using HTTP headers, and clients-side JavaScript cannot alter that. A proxy may be needed to work around strict CSPs. Also, the TrustedTypes API for security against cross-site scripting imposes restrictions on code injection.

Due to the default behavior of iframe and its cross origin access restrictions, intercepting and modifying the APIs is not trivial when the app uses WebRTC in an iframe. There are issues with inline options panel display on some apps.

The WebRTC APIs have evolved over a decade. We do not support the older style callback APIs, and only support the modern Promise-based. The webrtc-adapter shim layer used in many such projects already does the conversion.

Our software does not yet have cross-browser support. In particular, the end-user driven browser extension (usage option 1) is only available for Chromium-based browsers such as Chrome, Brave and Edge. Even in developer driven (usage option 2 or 3) installs, certain APIs have limited cross-browser compatibility, e.g., captureStream is not supported on Safari, and the MediaStream's inactive event is not triggered on FireFox.

## 6. SECURITY CONSIDERATION

Browser extensions are more powerful than web apps. A malicious extension can cause lasting damage, especially on end user privacy and data confidentiality. One way to limit access is to install the extension in a new browser profile, and to avoid doing regular web browsing in that profile. Inspecting the source code and network messages also helps but requires a deeper understanding of the software. Finally, users can install the unpacked browser extension locally in developer mode after updating the manifest file to remove any suspicious permissions. For developers, this approach is safer, not just for our software but also for other browser extensions.

Web developers should be wary of untrusted JavaScript code running in their web apps to avoid malwares affecting their users. End users should also avoid injecting untrusted code in the web apps they use. Our pre-built customization functions are open source, and thus, can be inspected and trusted, and the software loads those functions with an integrity check. However, if the end user or the web developer allows additional sources for downloading and installing more functions, a malicious actor may sneak in some malware code. Our software has a strict mode that can be configured to apply to such external sources. In strict mode, only a very limited set of essential programming constructs are available to the customization function, preventing any malicious behavior.

## 7. CONCLUSIONS AND FUTURE WORK

We have presented an endpoint driven software architecture and implementation to support customization of WebRTC and related APIs. This enables many novel features in existing and new web-based video conferencing, web collaboration, and audio/video related web apps. Our focus is on end user driven innovation, which is not common in this industry. We believe that unlocking the client-side WebRTC features in an app, away from the vendor driven service will promote great innovations in this space.

Endpoint driven features and applications [12] are inline with the open web philosophy and the end-to-end principle of the Internet. Unfortunately, this goes against the current trend of software-as-a-service web apps, which gravitate towards walled gardens with vendor lock-ins. We plan to continue exploring new ways to innovate in the endpoint against this ongoing trend.

From a web developer's perspective, the app development space is continuously evolving. However, app development often gets locked into a single framework due to poor architectural decisions. Soon, it becomes hard to break free, or to quickly try out new ideas. Our software enables web developers to break free in vanilla JavaScript independent of the framework baggage of the original web app. We also allow them to easily add additional functions for their users, and get early feedback.

Our software is still in its early stage. We mentioned a few limitations earlier. Additionally, we plan to work on cross browser compatibility, a graphical drag-drop style function editing (e.g., Scratch programming language), compatibility with popular video conferencing web apps, integration with virtual presence/digital trail, improvement and optimization of recording in conjunction with the backend systems, use of faster webgl or webgpu instead of 2d in canvas context, expansion of CPU category to include other resources such as network usage and memory, and support for audio track manipulation, e.g., using the web audio and related APIs. We will continue to add predefined functions for newer use cases. As more customization scenarios are explored or discovered, we will add more categories as well.

 

# Appendix A) ADDITIONAL USE CASES

We listed several motivational use cases in Section 2. Here, we describe some more customization examples. Most of these use cases mentioned earlier as well as here are already implemented as predefined functions in our software. These can be further customized by the end user if needed. This section is for informational purposes.

**Capture**: Ability to create images for video frames as webcam feed unlocks tremendous end user creativity. It allows popular features like virtual background, replacing face with avatar, decorating it with hats or glasses. It also enables integrating other image processing algorithms and emerging AI art, provided the client machine has enough real-time processing capacity.

For testing and debugging, we need the ability to suppress any frame generation, to dynamically change the resolution or framerate, or to show an analog (Fig. 9b) or digital clock to measure the media path latency subjectively.

It should be possible to send the webcam feed with some random or fixed color, or some random or user selected and locally uploaded picture, animated gif or video file, or logo or text overlay on the video. The controls variables could be used to parameterize, say, the file path, clock's time zone offset, picture's brightness or contrast, or the text content.

In terms of software implementation, replacing background with a static image file is almost the same as replacing it with an animated gif or a video file. This could show a more realistic virtual background in many situations.

Similar to the giphy or emoji features in popular text chat applications, or catchy audibles of Yahoo Messenger, user controls events can be used to temporarily replace or show as overlay, some animated gif with captions in video.

If the web app does not support screen share, or limits to a single share at a time, the user should be able to send her screen or window share as the webcam feed to emulate multiple screen shares.

If the end user has two or more webcams, she should be able to easily switch between them without depending on the app's ability, or to create a combined video feed from some or all of her webcams (Fig. 6), even if the web app allows only one video from each participant.

It should be possible to add text caption overlay using auto-transcription of the user's microphone audio. This can not only help other viewers to see what was said especially in a noisy environment, but also helps with note taking and accounting.

Due to the popularity of background detection in Capture use cases, we include a mediapipe-based body segmentation and face detection implementation. Besides blur, replace or removal of video background, it can also be used to blur and hide the foreground face and body for privacy (Fig. 1), or to replace the background with animated gif or video file.

The user should be able to select the picture or video file, or to change the behavior via the user controls event triggers.

The face detection algorithm is useful to highlight detected faces, or to auto-frame (zoom and pan) to the detected face, especially when the webcam feed is shown in a small size such as combined with the screen share video.

We also demonstrate receiving third-party live video, and sending that as the webcam feed in a call. We use the rtclite named stream service, but it should be possible to use any other such systems. As an example, this allows the end user to temporarily replace her conference webcam video with her mobile camera feed.

**Screen**: The functions described in Capture, can also be used in Screen. However, Screen is targeted for screen and window share scenarios, and may include audio. Common use cases such as adding a logo, or auto transcribed captions (Fig. 2) are more applicable to screen share presentations.

Overlaying webcam video in a small circle on top of screen or window share video is often used in online presentations and demonstrations (Fig. 2). This can further be enhanced using face detection to zoom and pan to show only the face, or to use body segmentation to make the presentation appear in the background while the foreground user is at a bottom corner (Fig. 11).

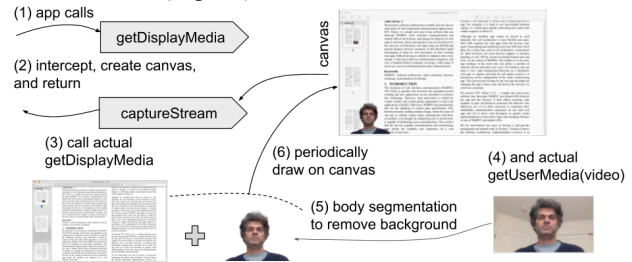

**Fig. 11** Modify Fig. 2 to remove background in webcam video.

Image processing can detect a rectangular region in a user's webcam video, such as a white board or a picture frame in the background, and draw the screen share content in that region, for an immersive presentation experience.

For privacy, unlike blur of face or body in a webcam video, we pixelate screen share content to avoid accidental leak of sensitive text. This applies to the content outside the active window, if full screen share is done, or to the whole content in some cases.

If the web app does not support sharing multiple windows, then the user should be able to select a subset rectangle on the full screen to share and send in a call, or to create one video feed that combines multiple selected windows' views.

**Record**: As mentioned earlier, conference recording is often an essential feature in existing video conferencing systems. However, such service driven recordings are often optimized for all users, and may not be the best for a user. We allow the end users to control what videos they would like to include in the recording, and to define the layout.

Creating a one-size-fits-all backend recording often restricts the number of user videos if content share is included, or limits the video resolution, or has rigid placement of videos in the combined layout. For example, the web app may allow the user in live call to select which other participants' webcams are shown, which may be different than the user videos included in the backend recording. This causes the viewing of recording to be a different experience than the user's live call. Another example is, if the end user wants to include only the shared presentation and the single active speaker in the recording, without the five other videos of the passive users just because they enabled their webcams in the call.

A common recording layout is to preserve the original web app layout of the videos as viewed by this end user, so that viewing the recording gives the right experience (Fig. 7). But many other different layouts are possible, e.g., a grid of all videos including local webcam preview even if the web app has the preview in a smaller size, or a smarter layout that accommodates webcam videos of varying aspect ratios while reducing the empty space. The end user may select to include only active videos, or all videos including disabled webcams to represent each participant as a box.

Some of the functions in Capture and Screen can be reused in Record, e.g., to place logo or text overlay with user names in the local recording, or to overlay the background removed webcam video of the speaker on the presentation video.

We use the JavaScript MediaRecorder API in the browser. The recorded content is available in the modern webm format by default, and can be downloaded by the end user. We also allow client side conversion to mp4 format using an external web assembly port of ffmpeg. However, this requires using the pcm audio codec in the recorder, and is not supported in FireFox.

We also show an example use case, where the end user can select and layout the videos in real-time for recording.

**Videos**: Unlike Record that allows layout changes in local recording, these customization functions allow changing the video conferencing layout visible in real-time at end point, and customizing the individual video element, e.g., to show the volume control or full screen button, even if the web app disabled it.

The user should be able to move or copy one or more video elements shown on the web app to a new external window. This could be done on direct user interaction such as click, or via controls event trigger. It could apply to any video element or only the active playing ones. The new window could display its own custom layout logic. With the help of a native application, it should be possible to show the windowless video elements, directly on the desktop screen with 3D effects or immersive user interface for a more appealing video conferencing experience.

Alternatively, one or more displayed videos could be combined in a single video, and shown as a floating video window always on top of other windows, using the modern browser's picture-in-picture web API.

**Stream**: The earlier function categories are enough to do a wide range of use cases, and we may not need to define any Stream customization. However, using an intermediate canvas to create the video frames is inefficient, and does not easily handle the audio track. On the other hand, a Stream function can manipulate a MediaStream object directly, with or without an intermedia canvas.

The local camera, microphone or display captured stream could be replaced by a locally uploaded video file, or an audio/video feed received from a third party service, e.g., to use a mobile or portable device as the media source in the web app opened on the laptop (Fig. 12). Since no canvas is used, these processes are more efficient than the similar Capture or Screen functions. When a locally uploaded video file is used as the camera and microphone stream, a separate user interface could display the local video player, and allow controls such as to pause or seek.

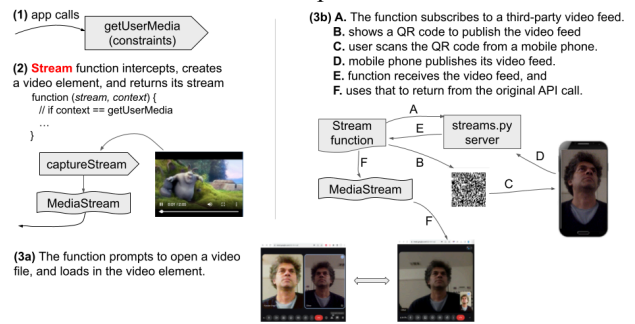

**Fig. 12** (a) locally uploaded video file as webcam, and (b) remote video feed from mobile camera as webcam in a Google Meet call.

For privacy or testing, the end user could force remove the microphone track from any locally captured stream, especially for screen share. Similarly, received streams could be intercepted to stop or remove the audio tracks, to avoid playing audio if needed. Received video could be temporarily paused or replaced with an external video, e.g., for force injecting or skipping a video ad.

Stream functions also intercept the media stream captured from an audio, video or canvas element. Users could disable such captures by the web app for privacy reasons, or display a warning text in the stream, or force overlay some logo or text label.

**Devices**: Most web video conferencing apps already have the ability to select the device for video input (camera), audio input (microphone) and audio output (speaker). But there are differences in how different platforms treat default devices. The selection can also be done in system settings, or in modern browser settings. Moreover, Windows also has separate system settings for default versus communication devices.

The user should be able to change or limit the list of devices exposed to the web app. This can help avoid problematic devices that crash or misbehave. It can also help privacy, by avoiding device finger printing by the web apps, e.g., by randomly changing device labels, hiding unique or explicit device names, or including dummy device names.

For testing and debugging the web app, the web developer could quickly hide certain device types, instead of having to physically detach those devices from the machine.

It allows the user to select devices only via system settings or other controls, and to expose only a single default device per type to the web apps.

**Media**: Ability to modify the capture constraints is crucial to web video conferencing apps. Usually this is done by the app behind the scenes, with no or little user input, such as quality versus performance tradeoff preference. But getting the right combination of framerate, resolution and bitrate for video is very tricky.

The `Media` functions can help developers experiment with different combinations without having to rebuild and redeploy the web app. Together with a `CPU` and/or `Stats` function, one can adjust the video capture constraints based on the network or processing load in real-time. If the web app captures webcam video at 30 fps and 720p causing extra CPU load, the end user can limit the app to, say, at most 10 fps and 320p to avoid heating up the laptop.

Privacy conscious users could disable certain media types, such as video input, if they join the call in audio-only mode. Similarly, they could remove audio input, if they plan to be a passive listener, such as in a seminar presentation. This avoids accidental audio recording by the web app. Note that modern machines often have a camera use indicator, but not a microphone use indicator.

Although the `Devices` function can modify the list of devices shown to the web app, the actual device selection is intercepted by the `Media` function. In particular, it can disregard the web apps preference, and select a specific audio or video input device. Even selecting front-facing vs. back cameras on mobile devices, and the ability to switch between the two is useful, independent of the web app.

**Session**: While a `Media` function deals with the capture constraints, a `Session` function is related to the encoding and decoding of media. WebRTC uses Session Description Protocol (SDP) to convey media codec parameters.

A `Session` function can print the complete session description during negotiation for debugging. It can disable a media such as video, or make it one way. It can select or prefer one codec over another such as to prefer H.264 over default VP8 for video, or prefer G.711 over default Opus for audio. For interoperability with media gateways, certain codec parameters can be configured, e.g., to disable stereo mode in audio, or to use certain packetization mode in H.264. It can manipulate the session at the low level, such as to add forward error correction (FEC) support, or remove negative acknowledgement support.

The function can impose bandwidth limits on both sender and receiver side. While the receiver bandwidth limit is set using a simple SDP attribute, the sender bandwidth is controlled using the encoding parameters.

The function can modify the `RTCRtpSender`'s encoding parameters, which includes attributes such as maximum bitrate or framerate for video, or clock rate and channels for audio. It can also affect `RTCRtpScriptTransform` that is used for end-to-end encryption of the media path, for privacy against any server side SFU (selective forwarding unit). All the endpoints must apply the function for this to work.

These functions together with triggers from resource usage can be used to automatically adjust the quality versus performance tradeoff in a video conference.

**Connect**: The configuration object in the constructor of `RTCPeerConnection` includes a list of external reflexive and relay server URLs and credentials. It can force the use of a media relay using the `iceTransportPolicy` attribute.

Strict enterprise policies [7] can be enforced using these controls, e.g., by injecting a specific media relay, forcing its use, and removing all other reflexive and relay servers.

In the past, Chrome allowed setting the network priority marking using DSCP (differentiated services code point) via the configuration, which the function can set as well.

**Network**: A `Network` function deals with the media path negotiation and establishment. In particular, it can intercept and modify the candidates used for ICE negotiations. Our predefined examples show how to remove candidates that are not applicable in a particular infrastructure or call flow scenario, e.g., those candidates that contain IPv6 or private IP addresses.

It can also change the list of candidates given to the application, e.g., by removing all but the specific relay candidates, to enforce strict enterprise policies mentioned previously, or to remove local IP addresses entirely for privacy, to avoid leaking them to the web app.

Note however that removing the candidate from the list given to the web app does not change the list of candidates already known to the browser, and the underlying WebRTC stack will continue to use them. For example, even if the function filters out a host candidate, the stack will continue to use that host IP address to send connectivity checks, and the actual media path may in fact get established with that host candidate. It affects only what the other endpoint sees in the signaling message, but does not affect what it sees in the media path as the source IP address.

**Stats**: When a `Stats` function is installed, the software intercepts almost all methods and events of the peer connection. The function can log all connection methods and events for testing and debugging purposes.

Additionally, the software periodically calls getStats on the peer connection, and passes the complete result object of type RTCStatsReport providing connection statistics. By default, this is done at a fixed interval that can be changed by the installed function. The customization function can use that to log the statistics data locally or send it to a monitoring service.

The function can also selectively display important metrics such as bitrate, delay, jitter, packets lost and available bitrate using graphs similar to that in webrtc-internals. Furthermore, the function can calculate a quality score based on primitive metrics, say, similar to the E-model for an objective mean-opinion-score (MOS) value.

Logging the difference between desired codec parameters and the actually used parameters is useful in detecting quality or load issues. For example, if the desired video resolution is 1080p, but the actual is less than 360p, it could indicate that either the high processing load or the low bandwidth has limited the encoded video resolution.

Since the Stats function can intercept most of the peer connection methods and events, it can also do what a Session, Network or Connect function can do, e.g., to change the video framerate of the encoder.

**Data**: The data channel intercept happens both in the RTCPeerConnection and the RTCDataChannel objects. The function can log all the messages sent and received on any data channel, or disable the data channel entirely, or replace it with some other signaling channel behind the scenes.

**Socket**: The WebSocket intercept is useful for web apps that use this type of connection for signaling. The intercept function can log all the messages sent and received on the socket, or display the signaling messages in a sequence diagram, or transparently encrypt and decrypt the messages to avoid intercept by network or browser devtools for reverse engineering, or change the target URL on the fly, or disable the socket entirely.

For testing and debugging robustness of the web app, a web developer can use this function to, say, force close the socket after a time or on controls, or to delay connection setup randomly, or to delay send or receive of messages, or consume the message sent to create a fake response as if received on the socket. Such testing strategies are often available as external tools, but using the customization function is a lot quicker to develop and test with.

**Request**: Standard JS Ajax requests and responses using the fetch or XMLHTTPRequest constructs are often used by the client web app to access server side APIs, and are sometimes used for WebRTC signaling as well. The intercept function can log all requests sent and responses received, or display the flow in a message sequence diagram, or transparently encrypt and decrypt the data sent or received, or compress and decompress the data, or change the target URL on the fly, or disable such requests entirely, or force the same-origin policy on all requests, or avoid sending cookies in requests and ignore them in responses, or disable the browser cache in web requests.

For testing and debugging robustness of the web app, a web developer can use this function to, say, randomly fake response with error on some requests, emulate success response with predefined content on others, automatically retry the request or fake response with error on timeout, or randomly delay sending or receiving data to emulate delay of server APIs.

Many of the use cases described above that are related to networking are useful for web developers for quick testing and debugging of their web apps, with or without WebRTC.

We described only a subset of use cases possible with our software. We will continue to evolve the software and the use cases in the future.

## Appendix B) FUNCTION PARAMETERS

Table II lists the various named parameters allowed in customization functions of different categories.

**Table II** List of named parameters.

| | |
|---|---|
| Capture | canvas, video, videos, screen, data, upload, segment, facedetect, respath, cleanup. |
| Screen | same as Capture, plus screen[1…n]. |
| Record | canvas, videos, states, data, usermedia, mediaconnect, noaudio, getstyle, cleanup. |
| Videos | video, open, data, getstyle. |
| Stream | stream, context. |
| Devices | devices. |
| Media | constraints, context. |
| Session | id, connection, session, context, data. |
| Connect | id, config/configuration, constraints, data. |
| Network | id, connection, candidate, context. |
| Stats | id, connection, type, name, args, data, videos, parsequery, query, plot, compress, send. |
| Data | id, connection, type, context, channel, args, data. |
| Socket | socket, type, context, argos, data. |
| Request | context, argos, xhr, resolve, reject, data. |
| CPU | details |

Some common parameters are allowed in all categories, and are often useful when the strict mode is enabled. These are global, console, location, and controls. All the parameters are described below.

*data*: an object that can be used by the function to save state across multiple invocations of the function, and is usually attached to the underlying context such as a video capture attempt in Capture, an RTCPeerConnection in Session, or a WebSocket in Socket. Some attributes in this object are set or interpreted by the software depending on the category, e.g., Capture and Screen set the start time of the capture,

Record interprets frameRate and mimeType to configure recording, Videos sets various state of the video element, and Request reuses the same data in both the request and its response for co-relation.

*context*: is a string representing the intercepted method or event, e.g., addIceCandidate (method) or icecandidate (event) for Network. A special value of construct or preconstruct is used to indicate the intercept of the object constructor, called after or just before.

*type*: a string, either "method" or "event", to indicate the intercept type, and is used together with the context string. For Data, an optional "RTCDataChannel." prefix is used to distinguish between the data channel versus peer connection intercept.

*name*: similar to context, representing a method or event name, and is used only in Stats for clarity, because it intercepts all methods and events of the peer connection.

*args*: an array of arguments passed to the intercepted method or event, and used together with context or name.

*id*: a string to uniquely identify a peer connection, which is assigned if the peer connection is intercepted.

*connection*: an associated RTCPeerConnection object.

*session*: RTCSessionDescription object representing an offer or answer of the intercepted method or event.

*candidate*: RTCIceCandidate object of the intercepted method or event.

*config/configuration*: the first parameter that was passed to the constructor of a peer connection.

*constraints*: either the object passed to the intercepted getUserMedia, getDisplayMedia or applyConstraints, or the second parameter that was passed to a peer connection constructor.

*channel*: an associated RTCDataChannel object.

*devices*: an array of MediaDeviceInfo objects.

*stream*: a MediaStream object containing zero or more tracks, or a MediaStreamTrack object for a streamless track.

*canvas*: an HTMLCanvasElement object used for drawing or creating the image frames for generated video.

*video*: an HTMLMediaElement for the captured webcam video in Capture or Screen, or an audio/video element on the web page.

*screen*: an HTMLVideoElement object representing captured screen or window share video.

*videos*: a list of HTMLMediaElement objects, representing captured video from all the webcams for Capture or Screen, or all the audio/video elements on the web page for Record.

*states*: an array of objects representing states of the video elements, for Record, in the same order as the associated elements in videos.

screen1, screen2, etc: If multiple such parameters are used, then the software allows multiple screen or window share; each object is an HTMLVideoElement representing one.

*cleanup*: if present, then the software invokes the function one last time during cleanup, and is useful in Capture, Screen and Record, where there is no other easy way to intercept or trigger cleanup in the function.

*details*: an array of objects containing performance metrics per CPU core.

*upload*: a convenience function to prompt file selection such as for virtual background or logo.

*respath*: path to the resources directory containing included image and video files, which are useful for demonstrations, and predefined functions. Note that the path is different based on the install options.

*controls*: an object to access, update or listen to a change in the shared variables of the Controls tab.

*segment* and *facedetect*: convenience functions to perform body segmentation and face detection, respectively, using the locally included or externally loaded mediapipe library.

*parsequery, query, plot, compress* and *send*: convenience functions used in Stats to interpret or parse peer connection statistics, display them graphically, or to compress and send them to an external monitoring service.

*usermedia, mediaconnect, noaudio*: are boolean flags that configure recording depending on which parameter is used, e.g., whether to include microphone captured audio, use audio from the peer connections instead of the audio/video elements, or disable audio in recording.

*socket*: an associated WebSocket object.

*xhr*: an associated XMLHTTPRequest object.

*resolve, reject*: are functions that can be called to return success or error immediately.

The following parameters are useful only in the strict mode. Note that the strict mode prevents access to even common objects such as window or navigator in the function.

*global*: an object representing global data, which can be used to share global variables across functions, and their calls. Unlike controls, there is no notification on change.

*console*: a wrapper to log to the JavaScript console.

*location*: the global location object for accessing web page URL, and its parameters.

*open*: to call the underlying window open function to open a new blank tab with a blob URL.

*getstyle*: to call getComputedStyle on a video element.

Note that the list of parameters, and which parameters are allowed in which category are obviously subject to change as the software evolves to include additional use cases.